**Employing both Gender and Emotion Cues to Enhance Speaker Identification Performance in Emotional Talking Environments**

Ismail Shahin

Electrical and Computer Engineering Department

University of Sharjah

P. O. Box  27272

Sharjah, United Arab Emirates

Tel: (971) 6 5050967

Fax: (971) 6 5050877

E-mail: ismail@sharjah.ac.ae

**Abstract**

Speaker recognition performance in emotional talking environments is not as high as it is in neutral talking environments. This work focuses on proposing, implementing, and evaluating a new approach to enhance the performance in emotional talking environments. The new proposed approach is based on identifying the unknown speaker using both his/her gender and emotion cues. Both Hidden Markov Models (HMMs) and Suprasegmental Hidden Markov Models (SPHMMs) have been used as classifiers in this work. This approach has been tested on our collected emotional speech database which is composed of six emotions. The results of this work show that speaker identification performance based on using both gender and emotion cues is higher than that based on using gender cues only, emotion cues only, and neither gender nor emotion cues by 7.22%, 4.45%, and 19.56%, respectively. This work also shows that the optimum speaker identification performance takes place when the classifiers are completely biased towards suprasegmental models and no impact of acoustic models in the emotional talking environments. The achieved average speaker identification performance based on the new proposed approach falls within 2.35% of that obtained in subjective evaluation by human judges.

## 1. Introduction and Prior Work

Speaker recognition systems are comprised of Speaker Identification (SI) and Speaker Verification (SV) systems. SI systems have the ability to determine the identity of an individual from a sample of one's voice. In such systems, the voice uttered by the unknown speaker is compared with model voices of all speakers in the speech data corpus. The comparison results are measures of the similarity from which the maximal quality is chosen. The applications of SI systems arise in criminal investigations to determine the suspected persons generated the voice recorded during the crime, calls to radio stations, local or other government authorities, monitoring people by their voices, and many other applications. SV systems possess the capability to accept or reject the identity of the claimed speaker. Such systems can be used in security control for confidential information areas, remote access to computers, and many other areas [1].

Speaker recognition using emotion cues is one of research fields for human-machine interaction or affective computing that has gained increasing attentions and concerns in recent years due to the wide variety of applications that benefit from such a new technology [2]. A major motivation comes from the desire to develop a human-machine interface that is more adaptive and reactive to a user's identity. The main function of the intelligent human-machine interaction is to give computers the ability of affective computing so that computers can identify the user for many different applications.

Speaker recognition field in emotional talking environments has many applications. This field is strongly related to the emotion recognition field. Some applications of the two fields appear in [3, 4, 5]:

1) Detecting the emotional state of the unknown speaker in telephone call center conversations and providing feedback to an operator or a supervisor for monitoring purposes.

2) Sorting voice mail messages according to the emotions expressed by callers.

3) Identifying the suspected persons who uttered emotional voice (*e.g.* sadness or anger) in emotional talking environments.

4) Text-To-Speech (TTS) communication-aid that can help expressing the correct emotion of the spoken text.

In literature, there are many studies that focus on speaker recognition field in emotional talking environments. Bao *et al*. [6] focused in one of their works on emotion attribute projection for speaker recognition on emotional speech. They used two methods to alleviate the emotion effects on speaker recognition on emotional speech. The first method is the emotion compensation method called Emotion Attribute Projection (EAP) which has been proved successful. The second method is the linear fusion of two subsystems, the Gaussian Mixture Model-Universal Background Model (GMM-UBM) based system and the Support Vector Machine (SVM) with EAP system. Li *et al*. [7] proposed an approach of speech emotion-state conversion to enhance speaker identification performance in the emotional talking environments. They tested their proposed approach using the Emotional Prosody Speech and Transcripts database [8]. In two of his recent studies [9, 10], Shahin focused on identifying the unknown speaker in emotional talking environments. In the first study [9], he proposed a two-stage approach based on both hidden Markov models (HMMs) and suprasegmental hidden Markov models (SPHMMs) as classifiers. He obtained an average speaker identification performance of 79.92% in a closed set with fifty speakers and six emotions [9]. In the second study [10], he focused on speaker identification in emotional talking environments based on each of HMMs, second-order circular hidden Markov models (CHMM2s), and SPHMMs. The achieved average speaker identification performance in a closed set with forty speakers and five emotions based on HMMs, CHMM2s, and SPHMMs is 61.40%, 66.40%, and 69.10%, respectively [10].

In general, human emotions are complicated phenomena, and many causes contribute to them. An entire definition of emotions must take into account the experience feeling of emotions, the processes that happen in the brain and nervous system and the observable expressive patterns of emotions [11]. There are many studies in the field of speech emotion recognition. Koolagudi and Krothapalli [12] proposed a two-stage speech emotion recognition based on speaking rate. At the first stage, eight emotions were categorized into three broad groups: active, normal, and passive. In the second stage, the three groups were classified into individual emotions based on vocal tract characteristics. The eights emotions were: neutral, anger, disgust, fear, happy, sadness, sarcastic, and surprise [12]. Lee and Narayanan [13] shed the light on recognizing emotions from spoken language. They used a combination of three sources of information for emotion recognition. The three sources are: acoustic, lexical, and discourse. Morrison *et al.* [14] aimed in one of their works to enhance the automatic emotional speech classification methods using ensemble or multi-classifier system (MCS) approaches. They also aimed to examine the differences in perceiving emotion in human speech that is derived from different methods of acquisition. Nwe *et al.* [15] proposed in one of their studies a text-independent method of speech emotion classification based on HMMs. Casale *et al.* [16] proposed a new feature vector that helps in enhancing the classification performance of emotional/stressful states of humans. The elements of such a feature vector are achieved from a feature subset selection method based on genetic algorithm. Wu *et al.* [17] proposed, implemented, and tested a new method called modulation spectral features (MSFs) for the automatic speech emotion recognition. These features were extracted from an auditory-inspired long-term spectro-temporal representation.

In the field of speech gender recognition, Lee and Narayanan [13] demonstrated that gender-specific emotion recognition performance is superior to the performance of both genders mixed. Ververidis and Kotropoulos [18] showed that the combined performance of separate male and female emotion recognition is higher than the performance of gender-independent emotion recognition. In one of their studies to enhance speech recognition performance through gender separation, Abdulla and Kasabov [19] separated the datasets based on gender to build gender-dependent HMM for each word. Their results showed significant enhancement of word recognition performance based on gender-dependent method over gender-independent method.

The contribution of this work is focused on proposing, implementing, and testing two distinct and separate approaches to enhance the performance of text-independent speaker identification in emotional talking environments. Specifically, this work focuses on improving the declined speaker identification performance in such talking environments using both speaker's gender and emotion cues. The first approach is based on a one-stage recognizer that uses SPHMMs as classifiers, while the second approach is based on a three-stage recognizer which employs both HMMs and SPHMMs as classifiers. The second approach is a continuation to one of our recent studies where emotion cues have been used to enhance speaker identification performance in emotional talking environments (two-stage approach) [9]. The goal of the proposed three-stage approach is to further improve speaker identification performance over the two-stage approach in such talking environments. Our collected speech database (CSD) has been used in the current work to separately evaluate the two proposed approaches. The CSD is composed of 6 emotions. These emotions are: neutral, angry, sad, happy, disgust, and fear.

The motivation of this work is that suspected speakers can be detected and tracked in emotional talking environments from both their gender and emotion cues. Another motivation is that speaker identification systems in such talking environments do not perform well as they do in neutral talking environments [10, 20]. Therefore, this work is focused on alleviating the low speaker identification performance in emotional talking environments.

The organization of this paper is as follows: Next section overviews the fundamentals of SPHMMs. Section 3 describes the collected speech database used to evaluate the two proposed approaches and the extraction of features. Section 4 is committed to discussing the two proposed approaches and the experiments.

Section 5 discusses the results obtained in this work. The conclusion of this work is given in Section 6.

## 2. Fundamentals of SPHMMs

SPHMMs were proposed, implemented, and evaluated by Shahin in three of his studies: speaker identification in emotional talking environments [10], speaking style authentication [21], and speaker identification under shouted talking condition [22]. SPHMMs have proven to be superior models over HMMs in these three studies.

SPHMMs are capable of summarizing several states of HMMs into a new state called suprasegmental state. Suprasegmental state possesses the capability to look at the observation sequence through a larger window. Such a state permits observations at rates appropriate for the situation of modeling emotional speech signals. Prosodic information can not be observed at a rate that is used for acoustic modeling. The prosodic features of a unit of emotional speech signal are called suprasegmental features because they affect all the segments of the unit speech signal. Prosodic events at the levels of phone, syllable, word, and utterance are modeled using suprasegmental states; on the other hand, acoustic events are represented using conventional hidden Markov states.

Polzin and Waibel [23] were able to combine and integrate prosodic information with acoustic information within HMMs as given by the following formula,

$$log\ P\left(\lambda^v, \Psi^v | O\right) = (1-\alpha) . log\ P\left(\lambda^v | O\right) + \alpha . log\ P\left(\Psi^v | O\right) \quad (1)$$

where $\alpha$ is a weighting factor. When:

$$\begin{cases} 0.5 > \alpha > 0 & \text{biased towards acoustic model} \\ 1 > \alpha > 0.5 & \text{biased towards prosodic model} \\ \alpha = 0 & \text{biased completely towards acoustic model and no effect of prosodic model} \\ \alpha = 0.5 & \text{not biased towards any model} \\ \alpha = 1 & \text{biased completely towards prosodic model and no impact of acoustic model} \end{cases} \quad (2)$$

$\lambda^v$ is the $v^{th}$ acoustic model, $\Psi^v$ is the $v^{th}$ SPHMM model, $O$ is the observation vector of an utterance, $P(\lambda^v | O)$ is the probability of the $v^{th}$ HMM model given the observation vector $O$, and $P(\Psi^v | O)$ is the probability of the $v^{th}$ SPHMM model given the observation vector $O$.

Equation (1) shows that leaving a suprasegmental state necessitates adding the log probability of this suprasegmental state given the relevant suprasegmental observations within the emotional speech signal to the log probability of the present acoustic model given the particular acoustic observations within the signal. More information about SPHMMs can be found in the references [10, 21, 22].

## 3. Collected Speech Database and the Extraction of Features

### 3.1 Collected Speech Database (CSD)

The speech database that has been used in this work was comprised of eight different sentences. These sentences were unbiased towards any emotion (*i.e.*, there was no correlation between any sentence and any emotion when uttered under the neutral state). These sentences are:

1) *He works five days a week.*
2) *The sun is shining.*
3) *The weather is fair.*
4) *The students study hard.*
5) *Assistant professors are looking for promotion.*
6) *University of Sharjah.*
7) *Electrical and Computer Engineering Department.*
8) *He has two sons and two daughters.*

A total of fifty (twenty five male and twenty five female students) untrained healthy adult native speakers of American English were separately asked to utter the eight sentences. The speakers were asked to portray each sentence nine times under each of the neutral, angry, sad, happy, disgust, and fear emotions. In this database, the speakers uttered the desired sentences naturally. These speakers were allowed to hear some recorded sentences before uttering the required database. The speakers were not allowed to practice generating such sentences under any emotion. The first four sentences of the CSD were used in the training phase, while the last four sentences were used in the evaluation phase (text-independent experiment).

This CSD was recorded in a clean environment that was not affected by a background noise. The CSD was captured by a speech acquisition board using a 16-bit linear coding A/D converter and sampled at a sampling rate of 16 kHz. The CSD was a wideband 16-bit per sample linear data. The signal samples were pre-emphasized and then segmented into frames of 16 ms each with 9 ms overlap between consecutive frames.

**3.2 Extraction of Features**

In this work, the features that represent the phonetic content of emotional speech signals in the CSD are called the Mel-Frequency Cepstral Coefficients (MFCCs). These coefficients have been widely used in the emotional speech and speaker recognition fields because of their superior performance over other features in the two fields and because of providing a high-level approximation of human auditory perception [24, 25, 26, 27].

MFCCs were computed with the help of a psycho acoustically motivated filter bank followed by a logarithmic compression and Discrete Cosine Transform. These coefficients can be computed as given in the following formula [28]:

$$C(n) = \sum_{m=1}^{M} \left\{ [log\ Y(m)] cos\left[ \frac{\pi n}{M}\left(m - \tfrac{1}{2}\right) \right] \right\} \qquad (3)$$

where *Y(m)* are the outputs of an *M*-channel filter bank.

A 16-dimension MFCC (8 static MFCCs and 8 delta MFCCs) feature analysis was used to form the observation vectors in each of HMMs and SPHMMs. An ergodic or fully connected HMM structure becomes more appropriate than a left-to-right (LTR) structure because every state in the ergodic structure can be reached in a single step from every other state. LTR structure is more appropriate than ergodic structure for isolated word recognition with a distinct HMM designed for each word in the vocabulary because time can be associated with model states in a fairly straightforward manner [29]. In this work, the number of conventional states, *N*, was nine and the number of suprasegmental states was three (each three conventional states were combined into one suprasegmental state) in SPHMMs and a continuous Gaussian mixture observation density was chosen for each model. Fig. 1 shows our adopted three-state ergodic SPHMMs which was obtained from a nine-state ergodic HMMs. The transition matrix, *A*, of such a structure can be expressed in terms of the positive coefficients $b_{ij}$ as,

$$A = \begin{bmatrix} b_{11} & b_{12} & b_{13} \\ b_{21} & b_{22} & b_{23} \\ b_{31} & b_{32} & b_{33} \end{bmatrix}$$

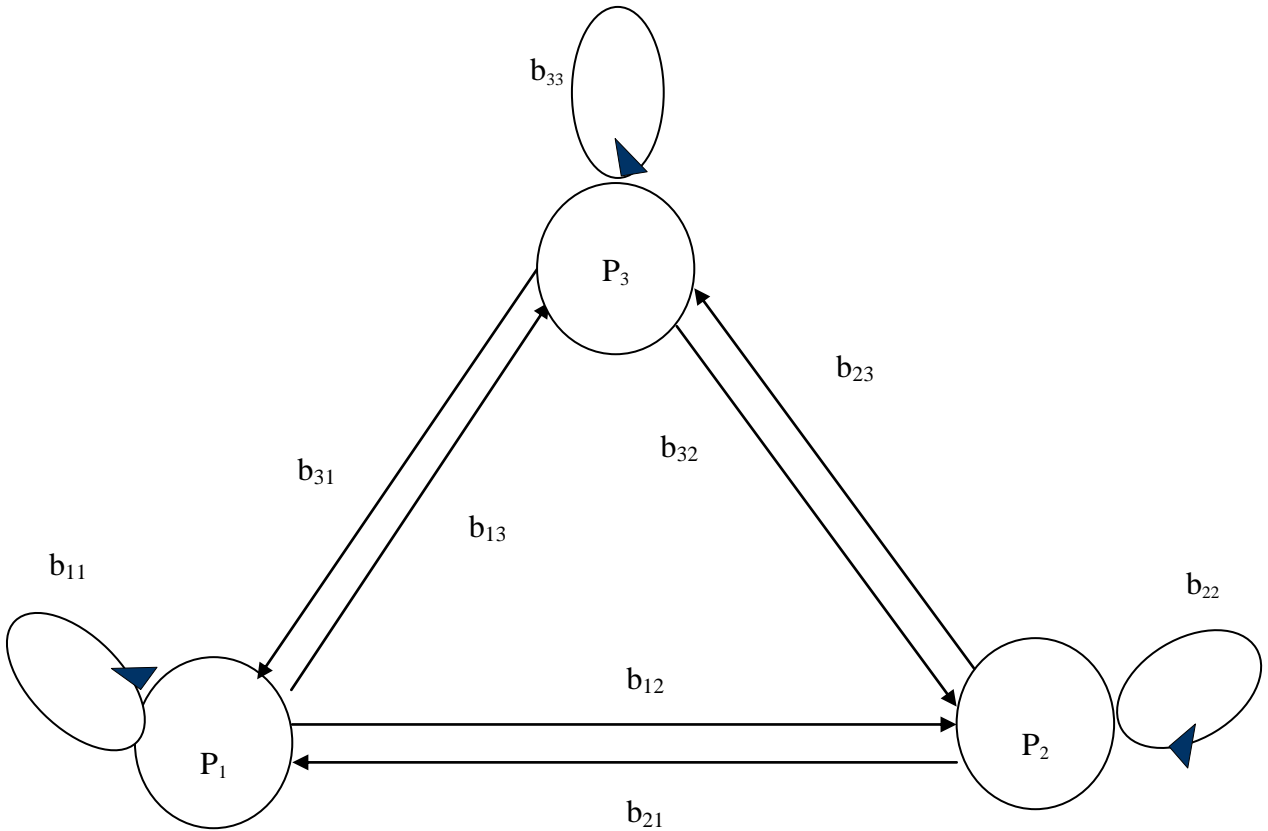

**Figure 1.** Three-state ergodic SPHMMs derived from a nine-state ergodic HMMs

# 4. Proposed Approaches of Speaker Identification in Emotional Talking Environments and the Experiments

## 4.1 Approach 1 (Simple Approach)

Given *n* speakers per gender where each speaker generates utterances under *m* emotions, there are *2nm* speaker models based on this approach. The probability of generating every utterance is computed based on SPHMMs (there are *2nm* probabilities), the model with the highest probability is chosen as the output of speaker identification as given in the following formula,

$$S^* = \arg\max_{2nm \geq u \geq 1} \left\{ P\left(O \mid \lambda^u, \Psi^u\right) \right\} \qquad (4)$$

where $S^*$ is the index of the identified speaker, $O$ is the observation vector or sequence that belongs to the unknown speaker with unknown gender and unknown emotion, and $P(O \mid \lambda^u, \Psi^u)$ is the probability of the observation sequence $O$ given the $u^{\text{th}}$ SPHMM model $(\lambda^u, \Psi^u)$.

In the training phase of SPHMMs, there are a total of 300 models (25 speakers × 2 genders × 6 emotions) where each model has been derived using nine utterances per sentence (the first four sentences of the database have been used in this phase). This phase is composed of 10800 utterances (25 speakers × 2 genders × 4 sentences × 9 utterances/sentence × 6 emotions). The training phase of SPHMMs is very similar to that of HMMs. In the training phase of SPHMMs, suprasegmental models are trained on top of acoustic models of HMMs.

In the evaluation (identification) phase, each speaker used nine utterances per sentence (the last four sentences of the database have been used in this phase to perform text-independent) under each emotion. The identification phase is composed of 10800 utterances (25 speakers × 2 genders × 4 sentences × 9 utterances/sentence × 6 emotions).

## 4.2 Approach 2 (Three-Stage or Hierarchical Approach)

Given *n* speakers per gender where each speaker talks in *m* emotions, the new proposed approach is composed of three cascaded stages as illustrated in Fig. 2. This figure shows that the new speaker identification approach is nothing but a three-stage recognizer that integrates and combines gender recognizer, emotion recognizer, and speaker recognizer into one system. The three stages are:

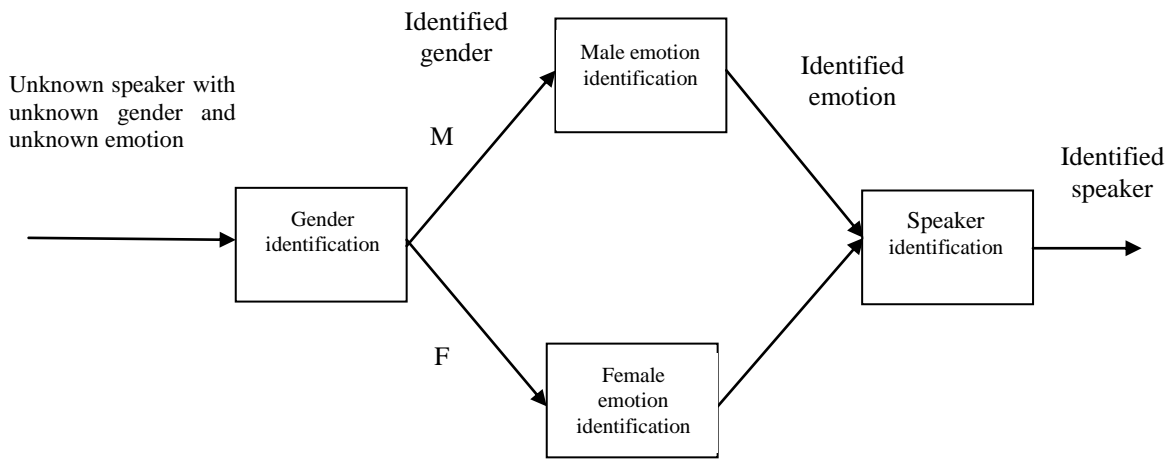

**Figure 2.** Block diagram of the overall proposed approach 2

**Gender Identification Stage**

The first stage of the proposed architecture is to identify the gender of the unknown speaker so as to make the output of this stage gender-dependent. The problem of differences in features for the two genders is well-known in the speaker recognition field [30]. Generally, automatic gender identification yields high performance with little effort since the output of this stage is the unknown speaker either a male or a female. Therefore, gender identification is a binary classification problem which is usually not very difficult stage.

In this stage, two probabilities per sentence are computed based on HMMs and the maximum probability is chosen as the identified gender as given in the following formula,

$$G^* = \arg\max_{2 \geq g \geq 1} \left\{ P\left(O \mid \Gamma^g\right) \right\} \tag{5}$$

where $G^*$ is the index of the identified gender (either $M$ or $F$), $\Gamma^g$ is the $g^{th}$ HMM gender model, and $P\left(O \mid \Gamma^g\right)$ is the probability of the observation sequence $O$ that belongs to the unknown gender given the $g^{th}$ HMM gender model.

In the training phase of this stage, HMM male gender model has been derived using the twenty five male speakers uttering all the first four sentences under all the emotions, while HMM female gender model has been obtained using the twenty five female speakers generating all the first four sentences under all the emotions. The total number of utterances used to construct each HMM gender model is 5400 (25 speakers × 4 sentences × 9 utterances/sentence × 6 emotions).

**Emotion Identification Stage**

Given that the gender of the unknown speaker was identified in the previous stage, the aim of this stage is to identify the unknown emotion that belongs to the unknown speaker. This stage is called gender-specific emotion identification. In this stage, *m* probabilities per gender are computed based on SPHMMs. The maximum probability is chosen as the identified emotion per gender as given in the following formula,

$$E^* = \arg \max_{m \geq e \geq 1} \left\{ P\left(O \middle| G^*, \lambda_E^e, \Psi_E^e\right) \right\} \qquad (6)$$

where $E^*$ is the index of the identified emotion, $\left(\lambda_E^e, \Psi_E^e\right)$ is the $e^{th}$ SPHMM emotion model, and $P\left(O \middle| G^*, \lambda_E^e, \Psi_E^e\right)$ is the probability of the observation sequence *O* that belongs to the unknown emotion given the $e^{th}$ SPHMM emotion model and the identified gender.

The $e^{th}$ SPHMM emotion model $\left(\lambda_E^e, \Psi_E^e\right)$ per gender has been derived in the training phase for every emotion using the twenty five speakers uttering all the first four sentences with a repetition of nine utterances/sentence. The total number of utterances used to construct each SPHMM emotion model per gender is 900 (25 speakers × 4 sentences × 9 utterances/sentence). The training phase of SPHMMs is very similar to the training phase of the conventional HMMs. In the training phase of SPHMMs, suprasegmental models are trained on top of acoustic models of HMMs.

**Speaker Identification Stage**

The last stage of the proposed architecture is to identify the unknown speaker given that both his/her gender and emotion were identified. This stage is gender-specific and emotion-specific speaker identification. In this stage, $n$ probabilities per gender per emotion are computed based on HMMs and the maximum probability is selected as the identified speaker per gender per emotion as given in the following formula,

$$S^* = \arg\max_{n \geq s \geq 1} \left\{ P\left(O \mid G^*, E^*, \Theta^S\right) \right\} \quad (7)$$

where $S^*$ is the index of the identified speaker, $\Theta^S$ is the $s^{th}$ HMM speaker model, and $P\left(O \mid G^*, E^*, \Theta^S\right)$ is the probability of the observation sequence $O$ that belongs to the unknown speaker given the $s^{th}$ HMM speaker model and both his/her identified gender and emotion. The $s^{th}$ HMM speaker model per gender per emotion has been constructed using nine utterances per sentence (the first four sentences of the database have been used in this phase). The total number of utterances used to build each gender-dependent and emotion-dependent HMM speaker model is 36 (4 sentences × 9 utterances/sentence). In this phase, neither the identified gender nor the identified emotion has been used in training the speaker identification module.

In the evaluation phase, each speaker of the CSD (closed set) used nine utterances per sentence (the last four sentences of the database have been used in this phase) under each emotion. The total number of utterances used in this phase is 10800 (25 speakers × 2 genders × 4 sentences × 9 utterances/sentence × 6 emotions).

# 5. Results and Discussion

In this work, two distinct approaches have been proposed, implemented, and tested on the CSD based on both HMMs and SPHMMs as classifiers. The value of the weighting factor ($\alpha$) is chosen to be equal to 0.5 to avoid biasing towards either acoustic or prosodic model. To the best of our knowledge, the two proposed approaches are the first attempt to employ both gender and emotion cues to identify the unknown speaker talking in emotional environments.

Table 1 summarizes speaker identification performance based on approach 1. It is evident from this table that the average speaker identification performance is low (78.25%). This is because approach 1 uses neither gender nor emotion cues to identify the unknown speaker. This table shows that the highest speaker identification performance occurs when female speakers speak in a neutral state (93%). The table also shows that the least speaker identification performance happens when female speakers speak in an angry emotion (70%).

Table 1

Speaker identification performance based on approach 1

| Emotion | Males (%) | Females (%) | Average (%) |
|---|---|---|---|
| Neutral | 92 | 93 | 92.5 |
| Anger | 71 | 70 | 70.5 |
| Sadness | 75 | 75 | 75 |
| Happiness | 78 | 79 | 78.5 |
| Disgust | 75 | 73 | 74 |
| Fear | 79 | 79 | 79 |

Based on the proposed approach 2, our automatic gender identification performance using HMMs is 96.87%. The achieved gender identification performance in this work is higher than that reported in some previous studies. Vogt and Andre [30] obtained gender identification performance of 90.26% using Berlin German database. Based on their study, Harb and Chen [31] reported gender identification performance of 92.00% in neutral talking environments. Gender-dependent emotion identification performance based on SPHMMs is given in Table 2. The average emotion identification performance of this table is 89.28%. This result is better than those obtained by:

i) Ververidis and Kotropoulos [18] who achieved male and female average emotion identification performance of 61.10% and 57.10%, respectively.

ii) Vogt and Andre [30] who obtained gender-dependent emotion identification performance of 86.00% using Berlin database.

Table 2

Gender-dependent emotion identification performance based on approach 2

| Emotion | Emotion identification performance (%) |
|---|---|
| Neutral | 97.8 |
| Anger | 84.5 |
| Sadness | 86.9 |
| Happiness | 90.1 |
| Disgust | 87.0 |
| Fear | 89.4 |

Tables 3 and 4 show male and female confusion matrices, respectively. One can notice the following from the two tables:

1) The most easily recognizable emotion is neutral, which is produced by female speakers (96%).

2) The least easily recognizable emotion is angry which is uttered by male speakers (81%).

3) Column 3 'Anger' of Table 3, for example, demonstrates that 4% of the utterances that were portrayed by male speakers in an angry emotion were evaluated as produced in a neutral state. This column shows that angry emotion has the highest confusion percentage with disgust emotion (10%). Therefore, angry emotion is highly confusable with disgust emotion. This column also demonstrates that angry emotion has no confusion with happy emotion.

Table 3

Male confusion matrix based on approach 2

|  | Percentage of confusion of a test emotion with the other emotions | | | | | |
|---|---|---|---|---|---|---|
| Emotion | Neutral (%) | Anger (%) | Sadness (%) | Happiness (%) | Disgust (%) | Fear (%) |
| Neutral | 95 | 4 | 1 | 6 | 2 | 3 |
| Anger | 0 | 81 | 5 | 2 | 7 | 3 |
| Sadness | 2 | 3 | 86 | 0 | 3 | 3 |
| Happiness | 3 | 0 | 0 | 88 | 2 | 1 |
| Disgust | 0 | 10 | 2 | 2 | 85 | 3 |
| Fear | 0 | 2 | 6 | 2 | 1 | 87 |

Table 4

Female confusion matrix based on approach 2

|  | Percentage of confusion of a test emotion with the other emotions | | | | | |
|---|---|---|---|---|---|---|
| Emotion | Neutral (%) | Anger (%) | Sadness (%) | Happiness (%) | Disgust (%) | Fear (%) |
| Neutral | 96 | 3 | 1 | 6 | 0 | 4 |
| Anger | 0 | 82 | 4 | 2 | 9 | 4 |
| Sadness | 2 | 3 | 85 | 2 | 3 | 4 |
| Happiness | 2 | 3 | 1 | 86 | 2 | 1 |
| Disgust | 0 | 7 | 2 | 2 | 84 | 1 |
| Fear | 0 | 2 | 7 | 2 | 2 | 86 |

Speaker identification performance of the overall proposed approach 2 when α = 0.5 and using the CSD is given in Table 5. This table gives an average speaker identification performance of 83.75%. Comparing Table 5 (average of 83.75%) with Table 1 (average of 78.25%), it is evident that approach 2 outperforms approach 1. This evidently shows that using both gender and emotion cues contributes in enhancing speaker identification performance in emotional talking environments.

Table 5

Speaker identification performance of the overall proposed approach 2 when α = 0.5 using the CSD

| Emotion | Males (%) | Females (%) | Average (%) |
|---|---|---|---|
| Neutral | 98 | 98 | 98 |
| Anger | 76 | 75 | 75.5 |
| Sadness | 84 | 83 | 83.5 |
| Happiness | 81 | 83 | 82 |
| Disgust | 84 | 82 | 83 |
| Fear | 80 | 81 | 80.5 |

A statistical significance test has been performed to show whether speaker identification performance differences (speaker identification performance based on approach 2 and that based on approach 1) are real or simply due to statistical fluctuations. The statistical significance test has been carried out based on the Student's $t$ distribution test. Based on the results achieved in this work, Tables 1 and 5 give: $\bar{x}_{approach\ 1} = 78.25$, $SD_{approach\ 1} = 7.64$, $\bar{x}_{approach\ 2} = 83.75$, $SD_{approach\ 2} = 7.55$. Using these values, the calculated $t$ value is $t_{app.\ 2,\ app.\ 1} = 3.618$. This calculated $t$ value is greater than the tabulated critical value at *0.05* significant level $t_{0.05} = 1.645$. Therefore, the conclusion that can be drawn in this experiment states that speaker identification based on the proposed approach 2 outperforms that based on the proposed approach 1. Hence, inserting both gender and emotion identification stages into speaker identification system in emotional talking environments significantly enhances speaker identification performance compared to that without inserting both stages.

It is apparent from the above tables and results that speaker identification performance based on the three-stage approach is high under the neutral state. This high performance is accredited to the fact that neutral state has almost the least confusion percentages with the other emotions as shown in Tables 3 and 4.

Combining gender, emotion, and speaker cues based on approach 2 into one system yields better speaker identification performance than those reported in some previous studies:

1) Gender-independent and emotion-independent speaker identification performance achieved by Shahin. In one of his studies [10], Shahin obtained an average speaker identification performance of 69.10% (in a closed set with 40 speakers and 5 emotions) in emotional talking environments based on SPHMMs. The average male and female speaker identification performance in such talking environments is 68.8% and 69.4%, respectively. Hence, it is evident that embedding both gender and emotion identification stages into gender-independent and emotion-independent speaker identification system significantly improves the identification performance compared to speaker identification recognizer without the two stages.

2) Gender-independent and emotion-dependent speaker identification performance reported by Shahin. In one of his recent studies [9], Shahin attained 79.92% as an average speaker identification performance (in a closed set with 50 speakers and 6 emotions) in emotional talking environments based on both HMMs and SPHMMs. The average male and female speaker identification performance in these talking environments is 79.50% and 80.33%, respectively. Therefore, it is apparent that adding a gender identification stage to the emotion-dependent speaker identification system significantly enhances speaker identification performance compared to emotion-dependent speaker identification recognizer without a gender identification stage.

3) Emotion-state conversion to enhance speaker identification performance in emotional talking environments. Li *et al.* [7] achieved 70.22% as a speaker identification performance using speech-state conversion method. This performance was obtained using the Emotional Prosody Speech and Transcripts database (EPD) with eight speakers talking in 14 emotions [8].

Seven more experiments have been separately conducted to evaluate approach 2. The seven experiments are:

i. Experiment 1: Gender-dependent, emotion-independent and text-independent speaker identification approach based on SPHMMs (25 speakers × 2 genders × 9 utterances/sentence × 6 emotions) has been carried out. This approach is composed of two cascaded stages: gender identification stage followed by speaker identification stage. A block diagram of the overall approach of this experiment is illustrated in Fig. 3. The training phase of the first stage is based on HMMs, while it is based on SPHMMs in the second stage.

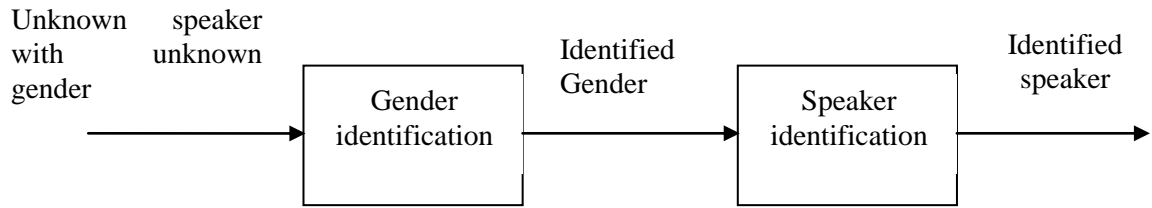

**Figure 3.** Block diagram of the overall approach of experiment 1

In this experiment, the achieved average speaker identification performance is 78.11%. The calculated $t_{app.\ 2,\ exp.\ 1} = 1.814$ which is greater than the tabulated critical value $t_{0.05} = 1.645$. This experiment leads to the conclusion that speaker identification based on gender-dependent and emotion-dependent approach is superior to that based on gender-dependent and emotion-independent approach.

ii. Experiment 2: Gender-independent, emotion-dependent and text-independent speaker identification approach based on SPHMMs (25 speakers/gender × 9 utterances/sentence × 6 emotions) has been performed. This approach consists of two cascaded stages: speaker identification stage preceded by emotion identification stage. Fig.4 demonstrates a block diagram of the overall approach of this experiment. SPHMMs have been used in the training phase of the emotion identification stage, while HMMs have been used in the training phase of the speaker identification stage.

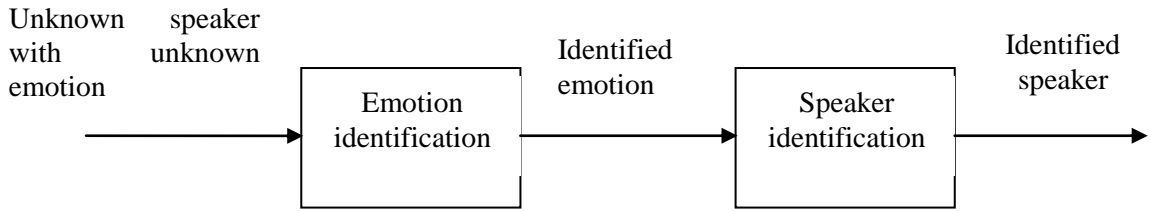

**Figure 4.** Block diagram of the overall approach of experiment 2

This experiment yields average speaker identification performance of 80.18%. The calculated $t_{app.\ 2,\ exp.\ 2} = 1.847$ which is higher than the tabulated critical value $t_{0.05} = 1.645$. It can be concluded from this experiment that speaker identification based on gender-dependent and emotion-dependent approach leads that based on gender-independent and emotion-dependent approach. Based on both Experiments 1 and 2, it is evident that emotion cues are more important than gender cues (relatively speaking) to identify speakers in emotional talking environments.

iii. Experiment 3: Gender-independent, emotion-independent and text-independent speaker identification approach based on SPHMMs (25 speakers/gender × 9 utterances/sentence × 6 emotions) has been conducted. The attained average speaker identification performance in this experiment is 70.05%. The calculated $t_{app.\ 2,\ exp.\ 3} = 2.432$ which is better than the tabulated critical value $t_{0.05} = 1.645$. Hence, speaker identification using both gender and emotion cues outperforms that using speaker cues only.

iv. Experiment 4: Approach 2 has been evaluated on a well-known emotional speech database called Emotional Prosody Speech and Transcripts database (EPD). This database was produced by the Linguistic Data Consortium (LDC) [8]. Such a corpus is composed of a limited number of speakers (three actors and five actresses) producing a series of semantically neutral utterances that consist of dates and numbers uttered in fifteen distinct emotions including the neutral state [8]. Only six emotions have been used in this experiment. These emotions are: neutral, hot anger, sadness, happiness, disgust, and panic. Table 6 shows speaker identification performance of the overall proposed approach 2 with α = 0.5 when this corpus has been used. The average speaker identification performance of this table is 81.17% which falls within 3.18% of that achieved in Table 5.

Table 6

Speaker identification performance of the overall proposed approach 2 when α = 0.5 using the EPD database

| Emotion | Males (%) | Females (%) | Average (%) |
|---|---|---|---|
| Neutral | 96 | 95 | 95.5 |
| Hot Anger | 73 | 74 | 73.5 |
| Sadness | 76 | 76 | 76 |
| Happiness | 84 | 85 | 84.5 |
| Disgust | 78 | 79 | 78.5 |
| Panic | 79 | 79 | 79 |

v. Experiment 5: The proposed three-stage approach has been assessed for distinct values of the weighting factor ($\alpha$). Figures 5 and 6 demonstrate average speaker identification performance based on this proposed approach for different values of $\alpha$ (0.0, 0.1, …, 0.9, 1.0) using the CSD and EPD, respectively. These two figures show that as the value of the weighting factor increases, average speaker identification performance in emotional talking environments (excluding the neutral state) enhances significantly. Therefore, it is apparent, based on the proposed three-stage architecture, that SPHMMs have more impact on speaker identification performance than HMMs in such talking environments. The two figures also demonstrate that the highest speaker identification performance happens when the classifiers are totally biased towards suprasegmental models and no influence of acoustic models (α = 1) in these talking environments.

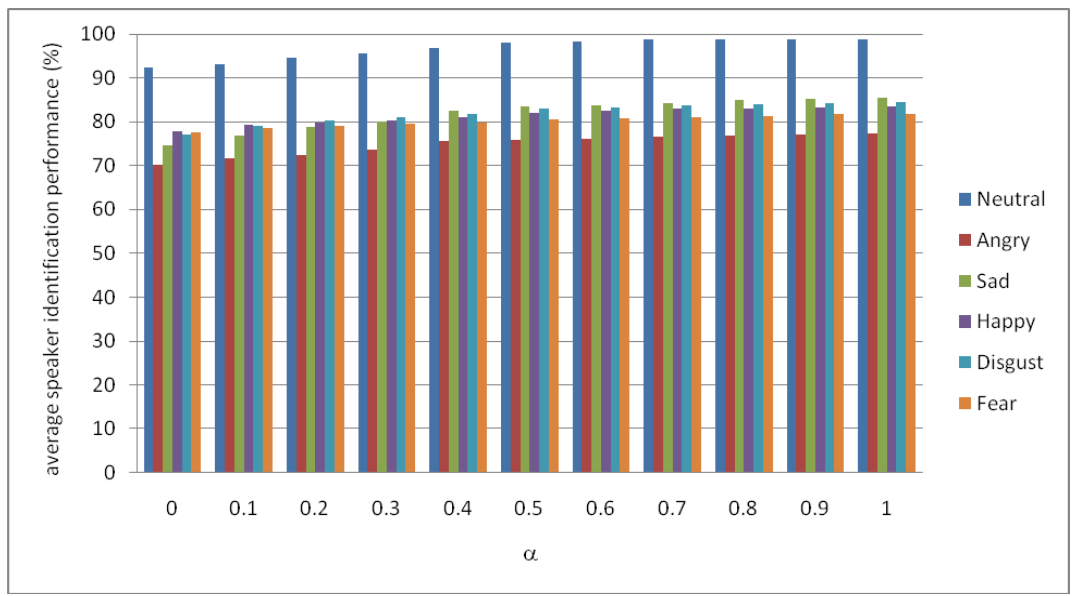

**Figure 5.** Average speaker identification performance (%) versus the weighting factor ($\alpha$) based on approach 2 using CSD

vi. Experiment 6: The proposed three-stage approach has been evaluated for the worst case scenario. Worst case scenario happens when each of the second (emotion) and third (speaker) recognizers receives false input from the previous stages (false identified gender from the first stage and false identified emotion from the second stage). The average speaker identification performance for the worst case scenario based on SPHMMs when α = 0.5 using the CSD is 78.51%. This average value is very close to that achieved using the one-stage approach (78.25%). The conclusion that can be drawn in this experiment is that speaker identification for the worst case scenario based on the three-stage approach performs almost the same as the one-stage approach.

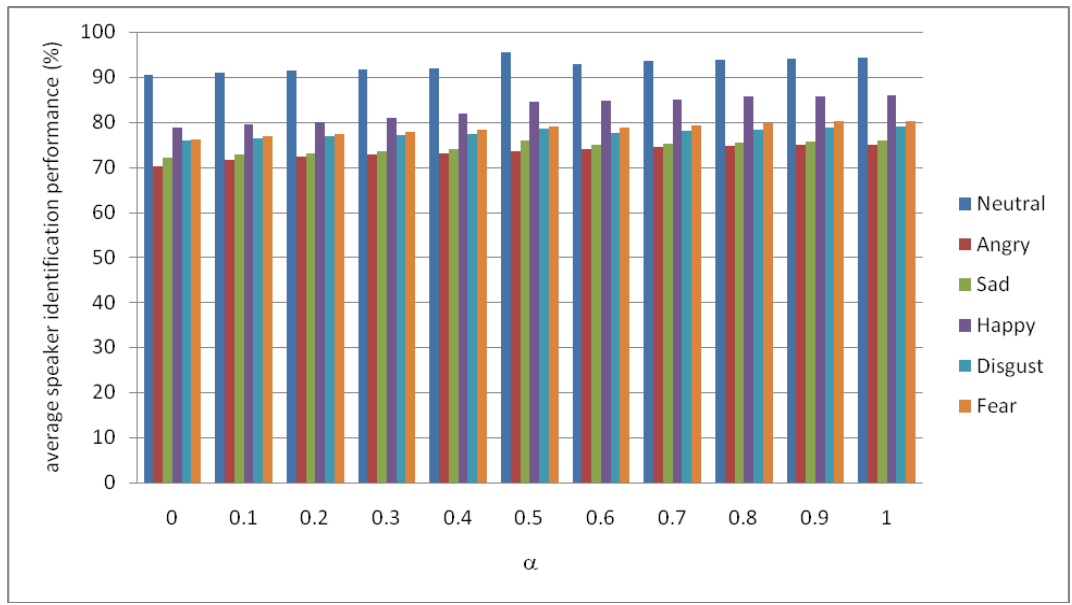

**Figure 6.** Average speaker identification performance (%) versus the weighting factor ($\alpha$) based on approach 2 using EPD

vii. Experiment 7: An informal subjective evaluation of approach 2 has been carried out using the CSD with ten (five male and five female students) nonprofessional listeners. These listeners were picked randomly from different ages of students. These students were not used in capturing the CSD. The ten students were not trained before conducting the evaluation experiment. A total of 1200 utterances (25 speakers per gender × the last 4 sentences of the database × 6 emotions) have been used in this evaluation. During the evaluation, the listeners are asked to answer three consecutive questions for every test utterance. The three questions are: identify the unknown gender, identify the unknown emotion, and finally identify the unknown speaker. The average gender, emotion, and speaker identification performance is 94.54%, 83.92%, and 81.83%, respectively. The average gender, emotion, and speaker identification performance obtained in this experiment falls, respectively, within 2.46%, 6.39%, and 2.35% of the achieved averages in the current work based on the proposed three-stage approach using the CSD.

## 6. Conclusion

In this work, a new approach based on using both gender and emotion cues has been proposed, implemented, and evaluated to enhance the low speaker identification performance in emotional talking environments. The current work shows that using both gender and emotion cues to identify the unknown speaker in such talking environments gives better speaker identification performance than using each of gender cues only, emotion cues only, and neither gender nor emotion cues. This is because the proposed three-stage approach possesses each of gender cues, emotion cues, and speaker cues all combined into one system. This work also shows that the highest speaker identification performance takes place when the classifiers are fully biased towards suprasegmental models and no impact of acoustic models in these talking environments. Finally, this work shows that the three consecutive recognizers perform almost the same as the single recognizer (one-stage approach) when each of the second and third recognizers of the three-stage approach receives false input from the preceding stages (false identified gender from the first stage and false identified emotion from the second stage).

There are some limitations in this work. First, the processing computations and the time consumed in the three-stage approach are greater than those in the one-stage approach. Second, the proposed three-stage approach requires all the emotions of the speaker to be available to the system in the training phase. Therefore, the proposed approach works only with a closed set system. Third, speaker identification performance using both gender and emotion cues based on approach 2 is limited. The performance of the overall proposed three-stage approach is the resultant of three performances. The reasons of the limitations are:

a) Gender identification performance is imperfect.

b) The emotion of the unknown speaker in the emotion identification stage is not 100% correctly identified.

c) The unknown speaker in the speaker identification stage is not perfectly identified.

For future work, we plan to thoroughly analyze the three-stage approach analytically to calculate the overall speaker identification performance and the performance in every single stage; we need to formulate a relationship between the overall performance and every stage performance. We also plan to propose new classifiers to improve speaker identification performance in emotional talking environments. In addition, we intend to evaluate our proposed three-stage approach on an unconstrained spontaneous emotional database.